# Spin Polarized Transport in Core Shell Nanowire of Silicon and Germanium

Bhupesh Bishnoi[1] and Bahniman Ghosh[1,2]

[1]DEPARTMENT OF ELECTRICAL ENGINEERING, INDIAN INSTITUTE OF TECHNOLOGY, KANPUR, 208016, INDIA

Email: bbishnoi@iitk.ac.in

[2]MICROELECTRONICS RESEARCH CENTER, 10100, BURNET ROAD, BLDG. 160, UNIVERSITY OF TEXAS AT AUSTIN, AUSTIN, TX, 78758, USA

Email: bghosh@utexas.edu

ABSTRACT

We investigate spin polarized electron transport in ultra-thin Si-Core/Ge-Shell and Ge-Core/Si-Shell nanowire system using semi-classical Monte Carlo simulation method. Depolarization of electron's spin occurs in nanowire mainly due to D'yakonov-Perel dephasing (DP-mechanism) and Elliott-Yafet dephasing (EY-mechanism). We studied the dependence of spin dephasing on ultra-thin silicon core diameter in Si-Core/Ge-Shell nanowire and germanium core diameter in Ge-Core/Si-Shell nanowire. Variation in spin dephasing length with varying core diameter ranging from 1 nm to 9 nm indicate that spin dephasing length increases with increase in Si-core diameter in Si-Core/Ge-Shell nanowire and spin dephasing length decreases with increase in Ge-core diameter in Ge-Core/Si-Shell nanowire. We then studied the variation in spin dephasing length with varying externally applied transverse electric field ranging from 20 kV/cm to 100 kV/cm. In the electric field dependence study we found that spin dephasing length is weakly dependent upon applied electric field. In the end, we studied the variation in spin dephasing length with varying temperature in the range of 4 K to 377 K. In this simulational study we found that for both Si-Core/Ge-Shell and Ge-Core/Si-Shell nanowire system spin dephasing length shows a strong dependence on temperature and spin dephasing length increases with decrease in temperature from room temperature range.

## I. INTRODUCTION

Spintronics has attracted a lot of research interest in the last decade with an aim to integrate magnetic materials and semiconductors. The motivation behind the spintronics technology is to use the spin of electrons as a tool to encode, store and transmit in an energy-efficient way digital data in modern electronics. [1-3] Spintronics technology based on semiconductor material can be used to integrate both logic processing, memory storage and communication device on a single chip. [4, 5] Hence, spintronics technology provides a path for multifunctional devices in classical sense. Also, spin is used in the domain of quantum computation. [6] There are basically three basic phenomena in the spintronics technology, (1) at the source information is stored in spin state and then injected. This process is known as spin injection. (2) Information is carried by electrons which experience scattering and hence the electron spin states relax along the channel. This is the method of spin dephasing. (3) Information is detected at the drain known as spin detection. [7, 8]An analysis described in our paper focuses on spin dephasing. Implementation of spin-based electronics in the mainstream semiconductor material such as Silicon (Si) and Germanium (Ge) is important challenge for the success of spintronics technology. [9-12]Other than conventional semiconductor nanowire structure, recently core-shell nanowire structure draws the attention of research community and offers larger carrier confinement due to their unique structure.

In the core-shell nanowire system band engineered dimension and material creates the band offsets in the energy band diagram and carriers are confined in the core material and effect of surface states is eliminated. Hence, effective conductance increases. In the Ge-core/Si-shell nanowire structure due to band alignment hole confinement potential has larger value and in the Si-core/Ge-shell nanowire structure electrons and holes confinement potential has smaller value. [13]

Another advantage of Ge-core/Si-shell nanowire structure is Ge-core has high intrinsic carrier mobilities and Si-shell can be used for chemical passivation of nanowire by oxidation process as formation of $SiO_2$ is well calibrated process in semiconductor technology and gives a stable higher quality surface passivation when compared to Ge based oxides. Ge-core/Si-shell nanowire channels based Field-effect Transistor (FETs) has shown better performance metrics compared to single-element material e.g. Ge, Si based nanowire FETs. [14] Ge-core/Si-shell nanowire exhibits significant quantum coherence effects accompanied by high hole mobility which provide an excellent one-dimensional nanoscale system. [15, 16] In core/shell nanowire structure scattering rates are reduced as electron transport is restricted only in the lateral direction and hence mean free path of carriers is several hundred nanometers at room temperature. [17] In the present work we studied spin polarized electronic transport in ultra-thin Si-Core/Ge-Shell and Ge-Core/Si-Shell nanowire system using a semi-classical Monte Carlo simulation approach. The effect of different parameter (total core–shell diameter, core diameter, electric field, temperature) on spin dephasing length is investigated. For Si-Core/Ge-Shell and Ge-Core/Si-Shell nanowire structure we perform simulation for the electric field in the ranges from 20 kV/cm to 100 kV/cm and the temperature ranges from 4 K to 373 K. Previously our group has studied spin relaxation under the influence of varying electric field and temperature in Silicon and Germanium nanowires. [18, 19] We also studied influence of variation of Germanium mole fraction on spin relaxation length in Silicon Germanium ($Si_{1-x}Ge_x$) and III-V group nanowires. [20, 21] In the rest of the paper spin polarized transport in ultra-thin Si-Core/Ge-Shell and Ge-Core/Si-Shell nanowires have been investigated and to the best of our knowledge no Monte-Carlo simulation work has been done in these very thin nanowires of up to few nanometers in diameter. Recently chemically synthesized Ge-core/Si-shell nanowire and their roughening and dislocation-mediated strain relaxation are reported. [14] In another study synthesis of Core-Shell Nano-needle Arrays of Black Ge is reported by Chueh *et al.*[22] In Ge-Si nanowire structure through magneto-transport study strong tunable spin-orbital coupling is investigated [15,23] In Ge-core/Si-shell nanowire structure helical hole states and strong spin-orbital interaction is reported by Kloeffel *et al.*[24] In another study of band offset configurations by doping in Si-Core/Ge-Shell nanowire, Amato *et al.*[25] show band offset variation in core-shell structure and experimental work has been done in Ge-core/Si-shell nanowire qubits for hole spin relaxation by Yongjie *et al.*[26] These studies show that core-shell nanowire structure is strong contender for future spintronics devices.

## II. MODEL

Jacoboni *et al*. first present Monte-Carlo simulation model for charge transport [27] and then spin transport in a quantum wire is studied by Pramanik *et al.* using a semi-classical approach. [28] Later Saikin *et al*. utilize kinetic transport equation, drift-diffusion and Monte-Carlo simulation approaches to summarize spin dynamics and transport in semiconductor structures. [29] Monte Carlo simulation approaches are extensively used for transport in nanowire [30-32] and modeling of spin FETs. [33] Recently modeling and analysis of Si-core/SiGe-shell nanowire has been done by Tang *et al*. [34] and shows greater quantum mechanical effect and carrier confinement in core-shell structure compared to semiconductor alloy nanowire. Hence core-shell structure shows promising thermoelectric-related transport property. In this article, we take core-shell structure with silicon as core material in Si-Core/Ge-Shell nanowire, germanium as core material in Ge-Core/Si-Shell nanowire and studied spin transport property by using semi-classical Monte-Carlo simulation. In the present article we only discuss key modification and necessary feature of Monte-Carlo simulation. For simplicity we approximate the cross section of core-shell nanowire as square as shown in Fig.1.The coordinate system is chosen where X is along the length of the core-shell structure, Y is along the width of the core-shell structure and Z is along the thickness of the core-shell structure. In the 1-D system electrons are restricted in the Y direction and the Z direction. Both the materials e.g. Silicon and Germanium possess bulk inversion symmetry. [35, 36] Hence, bulk inversion asymmetry or Dresselhaus spin-orbital interaction is absent in both Si and Ge. Therefore Dresselhaus spin-orbital interaction is also absent in Si-Core/Ge-Shell and Ge-Core/Si-Shell nanowire system. The transverse electric field which breaks structural inversion asymmetry results in Rasbha spin orbit interaction. Spin polarization hence occurs in core-shell system along the channel due to Rasbha spin orbit coupling via D'yakonov-Perel (DP) relaxation. [37, 38]

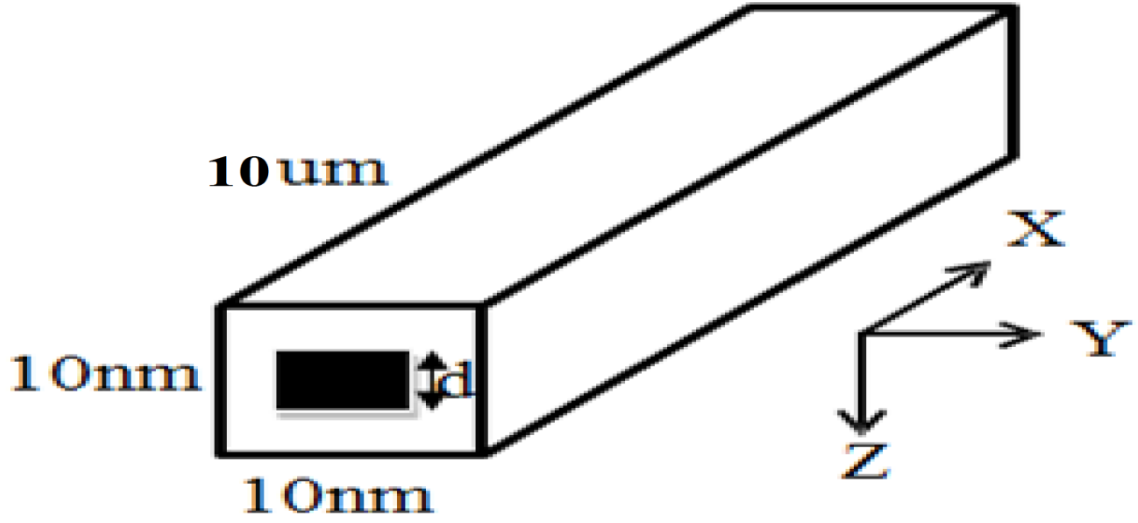

Fig.1. Core-Shell nanowire structure with core diameter of 'd'

During free flight in which no scattering occurs, the temporal evolution of spin is given by following equation

$$\frac{d\vec{S}}{dt} = \vec{\Omega}_{\text{eff}} \times \vec{S} \qquad (1)$$

Precession vector $\vec{\Omega}_{\text{eff}}$ has only Rasbha component which is given by

$$\Omega_R(k_x)^{1D} = -\frac{2\eta k_x}{\hbar}\hat{j} \qquad (2)$$

Where $\eta$ is Rasbha coefficient given by, [39]

$$\eta = \frac{h}{2m^*}\frac{\Delta_{so}}{E_g}\frac{(2E_g + \Delta_{so})}{(E_g + \Delta_{so})(3E_g + 2\Delta_{so})}e\vec{E} \qquad (3)$$

Where $\Delta_{so}$ is the spin orbit splitting, $e$ is electron charge, $\vec{E}$ is transverse electric field, $E_g$ is the band gap and $m^*$ is effective mass of electron.

Using Eq. (2) in Eq. (1), with spin vector $\vec{S}$ expressed as

$$\vec{S} = \vec{S}_x\hat{i} + \vec{S}_y\hat{j} + \vec{S}_z\hat{k} \qquad (4)$$

We get components of spin vector $\vec{S}$, expressed as

$$\frac{d\vec{S}_x}{dt} = -\frac{2}{\hbar}\eta k_x \vec{S}_z \qquad (5)$$

$$\frac{d\vec{S}_y}{dt} = 0 \qquad (6)$$

$$\frac{d\vec{S}_z}{dt} = -\frac{2}{\hbar}\eta k_x \vec{S}_x \qquad (7)$$

Sudden spin flipping happens due to Elliott-Yafet (EY) relaxation mechanism [40] and spin-flip scattering rate is given by,

$$\frac{1}{\tau_s^{EY}} = A\left(\frac{k_b T}{E_g}\right)^2 \alpha^2 \left(\frac{1-\alpha/2}{1-\alpha/3}\right)\frac{1}{\tau_p} \qquad (8)$$

Where $E_g$ is the band-gap, $\alpha = (\Delta/E_g + \Delta_{so})$, where $\Delta_{so}$ is spin-orbital splitting and $\tau_p$ is momentum relaxation time. $A$ is dimensionless constant and varies between 2 and 6. We have considered $A = 4$. The non-parabolicity approximation of the band is accounted by following energy-wave vector relation: [41]

$$\epsilon(1 + \alpha\epsilon) = \frac{\hbar^2 k^2}{2m^*} \qquad (9)$$

Where $\alpha$, is a non-parabolicity parameter and is given by the following expression,

$$\alpha = \frac{1}{E_g}\left(1 - \frac{m^*}{m_0}\right)^2 \qquad (10)$$

We have assumed that phonons of core region are confined in core and similar assumption is considered for shell region also. In our simulation, we have considered surface roughness scattering, optical phonon scattering, spin flip scattering, ionized impurity scattering and acoustic phonon scattering mechanism. [42-46]

## III. SIMULATION FRAME WORK

Spin polarized electronic transport in the core/shell nanowire structure is simulated based on models described in the section II. The total core-shell nanowire system cross section area is 10 nm × 10 nm and length is 10 um. In the core/shell nanowire structure we applied transverse electric field of 100 kV/cm and this effective electric field is responsible for Rasbha spin orbit coupling as it acts as symmetry breaking field. In our model we have taken four lower valleys and the energy levels of subbands are calculated by method of finite differences.[45] Reasonable value of driving electric field is chosen to restrict the majority of electrons in four subbands. The cross-sectional dimensions of the nanowire are also considerably small, such that the higher subbands will be at high energy levels and thus can be safely considered to be unpopulated.

The material parameters for Si and Ge are taken from Jacoboni *et al.* [27] and are given in Table I.

TABLE I

| Simulation parameter for Si and Ge | Ge | Si |
|---|---|---|
| Longitudinal mass | $1.58m_0$ | $0.98m_0$ |
| Transverse mass | $0.081m_0$ | $0.19m_0$ |
| Bandgap (eV) at 300K | 0.66 | 1.12 |
| Density (g/cm$^3$) | 5.32 | 2.33 |
| Speed of sound (cm/s) | $5.4 \times 10^5$ | $9.0 \times 10^5$ |
| Static dielectric constant | 16 | 11.8 |
| Non-parabolicity factor (eV$^{-1}$) | 0.65 | 0.5 |
| Acoustic phonon deformation potential (eV) | 11.0 | 9.0 |
| Optical phonon coupling constant (eV/cm) | $3 \times 10^8$ | $2 \times 10^8$ |
| Optical phonon temperature (K) | 320 | 685 |
| Lande g-factor | 1.563 | 2.0 |
| Spin Orbit Splitting (eV) | 0.29 | 0.044 |

The simulations are run for 1 million time steps with time steps of 0.2 femto-seconds so that electrons will reach their steady state condition. Data is recorded for last 50,000 steps and for all the components of spin vector $\vec{S}_x, \vec{S}_y$ and $\vec{S}_z$ ensemble average is computed according to the following expression, [47]

$$<\vec{S_i}>(x) = \frac{\sum_{t=t1}^{t=T} \sum_{n=1}^{n_x(x,t)} s_{n,i}(t)}{\sum_{t=t1}^{t=T} n_x(x,t)} \qquad (11)$$

Where $s_{n,i}(t)$ represents the value of the $i^{th}$ spin component of the $n^{th}$ electron at time $'t'$, $n_x(x,t)$ is total number of electron in a grid of distance $\Delta x$ around position $x$ at time $'t'$, $'t1'$ is the time where we start recording the data, $'i'$ denotes the $x, y$ and $z$ components, $'T'$ is the end time. The magnitude of ensemble averaged spin vector $|<\vec{S}>(x,T)|$ is given by,

$$|<\vec{S}>(x,T)| = \sqrt{<\vec{S}_x>^2 + <\vec{S}_y>^2 + <\vec{S}_z>^2} \qquad (12)$$

Spin dephasing length is expressed by distance travelled by electron from the point of injection to the point where $|\vec{S}|$ falls to one upon exponential times of its initial value at injection point. Z-spin polarized electrons are injected with 100% initial polarization hence initial value of $|<\vec{S}>(x,T)|$ is one.

## IV. RESULTS

A. Variation in magnitude of ensemble averaged spin vector $|S|$ and spin dephasing length variation with core diameter in Si-Core/Ge-Shell and Ge-Core/Si-Shell nanowire:

Here we investigate the spin dephasing length for various core diameters with the total core-shell diameter fixed for transverse effective electric field of 100kV/cm at room temperature (300K). From Figures 2-5, we show variation in magnitude of ensemble averaged spin vector with core diameter in Si-Core/Ge-Shell and Ge-Core/Si-Shell nanowire.

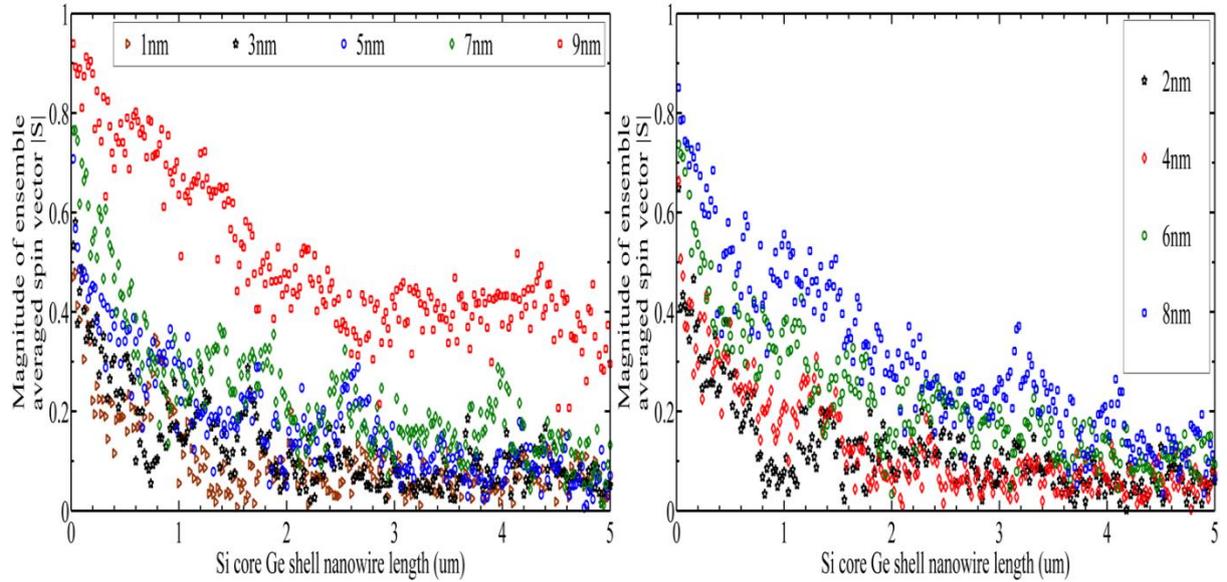

Fig.2. Variation in magnitude of ensemble averaged spin vector $|S|$ with varying core diameter in Si-Core/Ge shell nanowire

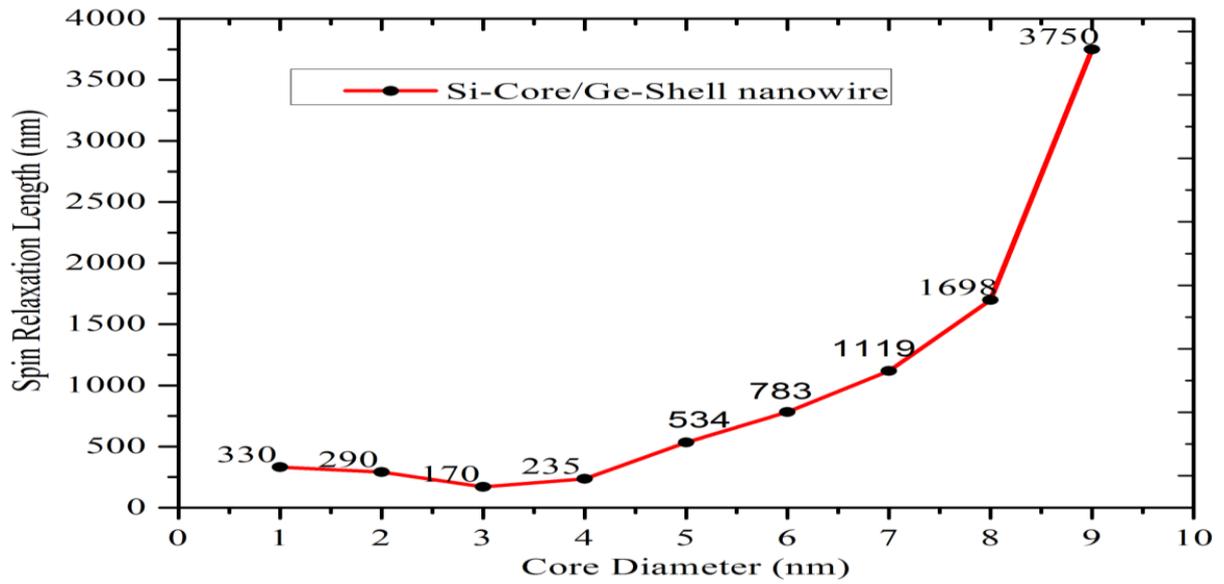

Fig.3. Variation in spin dephasing length with varying core diameter in Si-Core/Ge-Shell nanowire

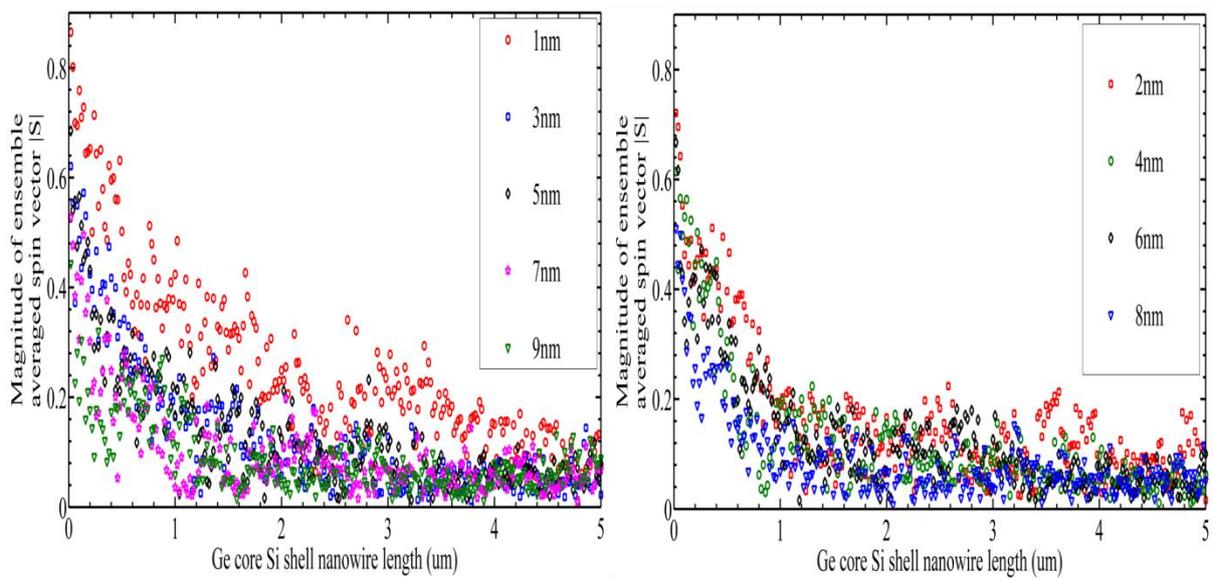

Fig.4. Variation in magnitude of ensemble averaged spin vector $|S|$ with varying core diameter in Ge-Core/Si-Shell nanowire

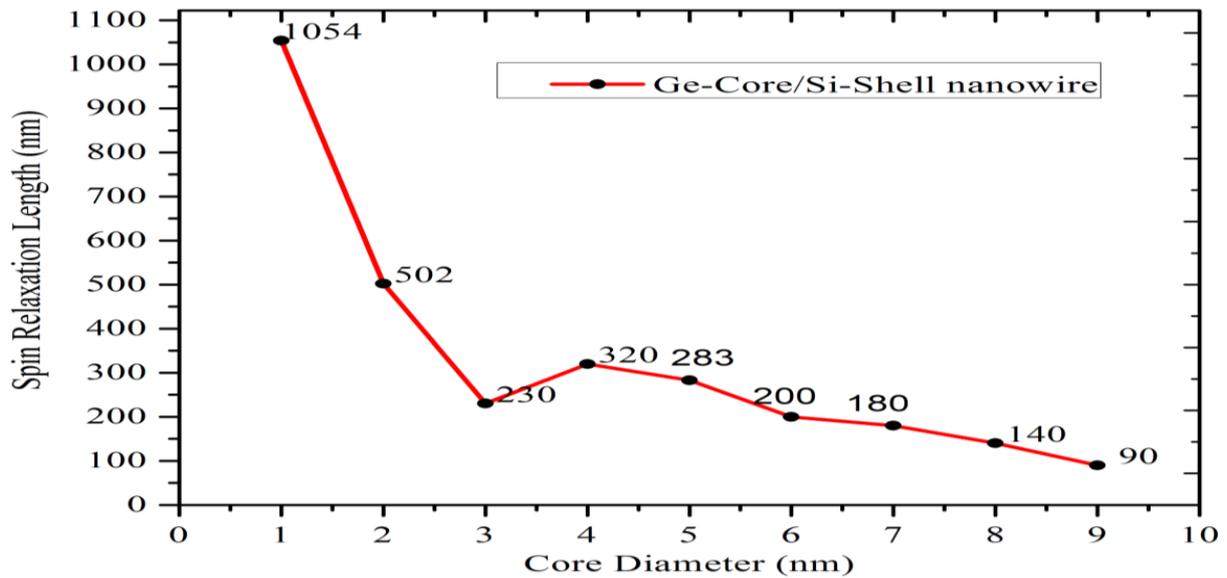

Fig.5. Variation in spin dephasing length with varying core diameter in Ge-Core/Si-Shell nanowire

In Figures 3 and 5, we show the variation of spin dephasing length with varying core diameter and it clearly indicates that spin dephasing length decreases with increase in Ge-core diameter in Ge-Core/Si-Shell nanowire and spin dephasing length increases with increase in Si-core diameter in Si-Core/Ge-Shell nanowire. Difference primarily lies in different band profiles obtained by different material combinations in Core/shell structure and hence carriers are confined according to Core/shell structure. The dimensions of the core-shell nanowire system govern the quantum confinement effects. The thickness of core and shell region is critical in determination of band structure and subband energy levels. In our Si-Core/Ge-Shell and Ge-Core/Si-Shell system there is a band offset at the interface between core and shell as shown in the Figure 6. When band-gap of core region is smaller than that of shell region, core acts as a potential well. When the core diameter is very small, the effect of quantum confinement is more significant. The energy of carriers confined in the core region becomes very high.

Table II shows the Spin dephasing length for various core diameters of Si-Core/Ge-Shell and Ge-Core/Si-Shell nanowire

TABLE II

| Core Diameter (nm) | Spin Dephasing Length(nm) Ge-Core/Si-Shell nanowire | Spin Dephasing Length(nm) Si-Core/Ge-Shell nanowire |
|---|---|---|
| 1 | 1054 | 330 |
| 2 | 502 | 290 |
| 3 | 230 | 170 |
| 4 | 320 | 235 |
| 5 | 283 | 534 |
| 6 | 200 | 783 |
| 7 | 180 | 1119 |
| 8 | 140 | 1698 |
| 9 | 90 | 3750 |

The subband energy levels exceed the band offset at interface. Most of the carriers have higher probability of being found in shell region. However, as we increase the core diameter keeping total nanowire diameter constant, the subband energy levels get lowered due to reduced quantum confinement effects. In this case, most of the carriers are confined to the core region. This is reflected in Figures 3 to 5 showing the variation of spin dephasing length with core diameter. In our simulations, we observed that for value of core diameter of 1 nm, in Ge-Core/Si-Shell nanowire almost all carriers were confined to shell region. Shell is composed of Silicon which exhibits higher spin dephasing length than Germanium and so spin dephasing length is high for these values of core diameter and as core diameter increases from this value confinement is reduced and carriers start distributing in core region and spin dephasing length decreases. For core diameters above 3 nm, spin dephasing length decreases by 80% and most of the carriers were trapped in the core region. Core region is composed of Germanium which exhibits lower spin dephasing length compared to Silicon and so the spin dephasing length is quite low in that region. On the other hand in Si-Core/Ge-Shell nanowire increment in Si-Core diameter increases the spin dephasing length as more and more carriers are confined in Si-Core regions and silicon has high value of spin dephasing length compared to germanium. Above 8 nm of Si-Core diameter almost all the carriers are confined in Si-Core region and this gives rise to higher spin dephasing length. Whereas for the intermediate values, it was observed that the distribution of carriers was a bit more uniform and the value of spin dephasing length is in mid-range.

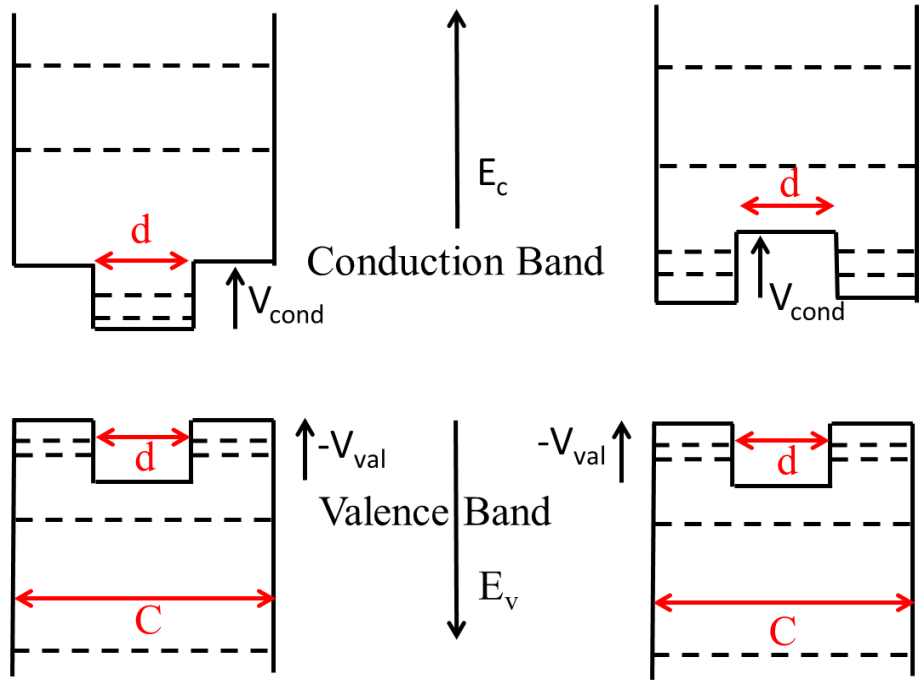

Fig.6. Band diagram of (a) Ge-Core/Si-Shell nanowire and (b) Si-Core/Ge-Shell nanowire

B. Variation in magnitude of ensemble averaged spin vector $|S|$ and spin dephasing length variation with applied electric field in Si-Core/Ge-Shell and Ge-Core/Si-Shell nanowire:

Figures 7-10 show the variation in magnitude of ensemble averaged spin vector $|S|$ and Spin dephasing length variation with applied electric field in core diameter of 4 nm for Si-Core/Ge-Shell and Ge-Core/Si-Shell nanowire. We observed that spin dephasing length depends weakly on electric field and the variation in itself is non-monotonic in nature. When increment in drift velocity dominates the increase in scattering rates, the spin penetrates deeper in the channel, giving us higher values of spin dephasing lengths. However, when scattering rates dominate drift velocity, spin dephasing is faster. The overall effect is governed by the prevailing effect of the scattering rate and drift velocity. Scattering rates saturate at higher field value. It remains nearly constant with only slight variation. Driving electric fields are selected so that drift velocity does not get saturated. Thus in high electric field region spin dephases slower due to dominance of drift velocity over scattering rates.

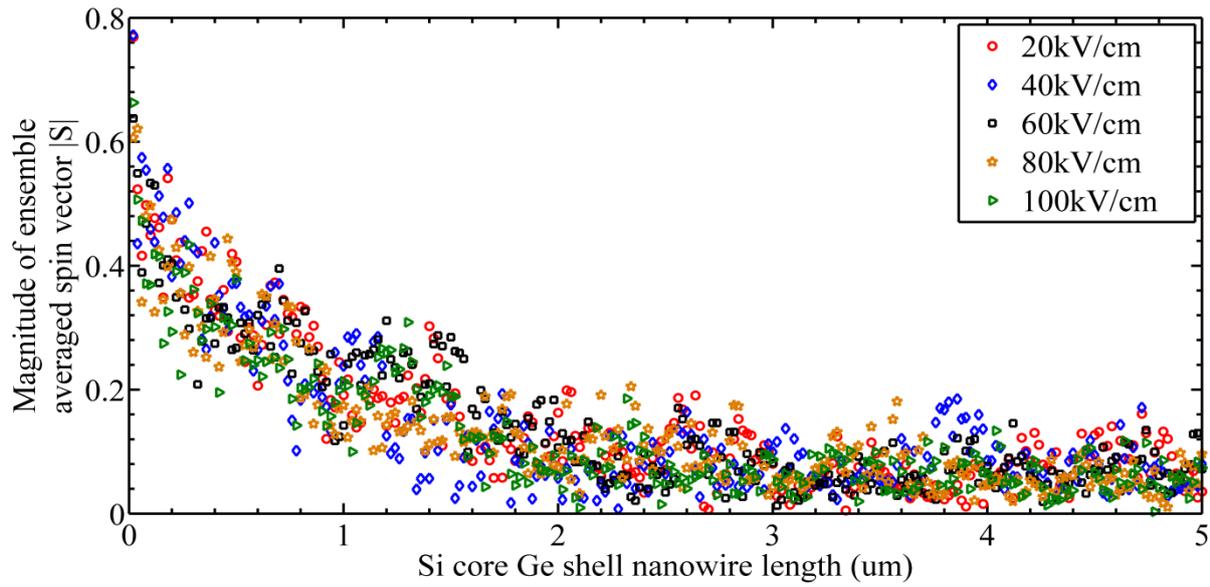

Fig.7. Variation in magnitude of ensemble averaged spin vector $|S|$ with varying electrical field in Si-Core/Ge Shell nanowire

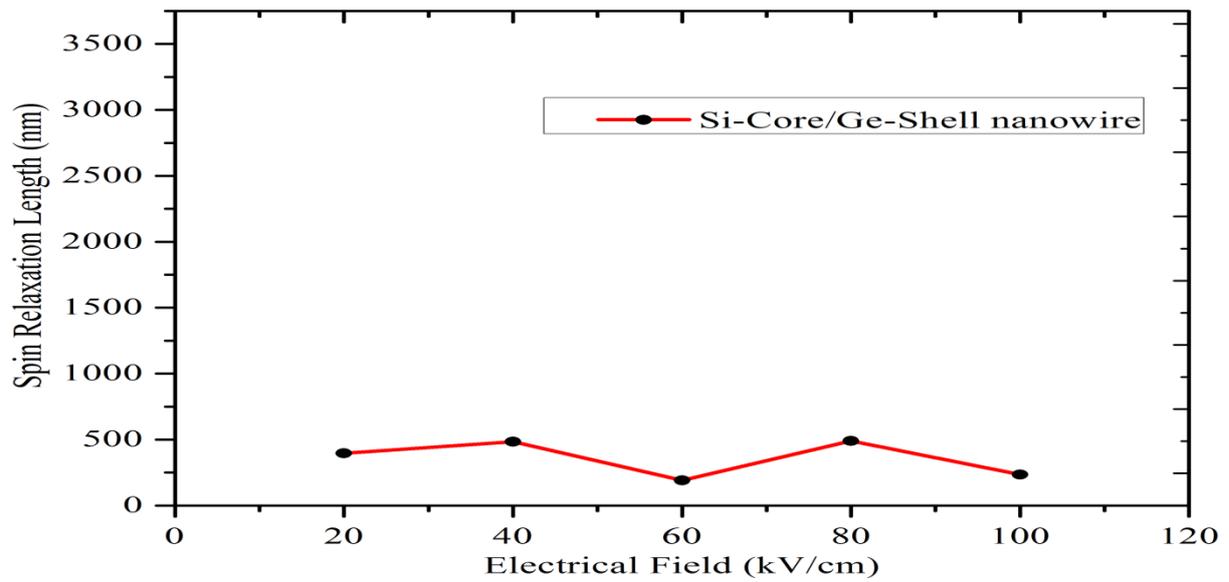

Fig.8. Variation in spin dephasing length with varying electrical field in Si-Core/Ge-Shell nanowire

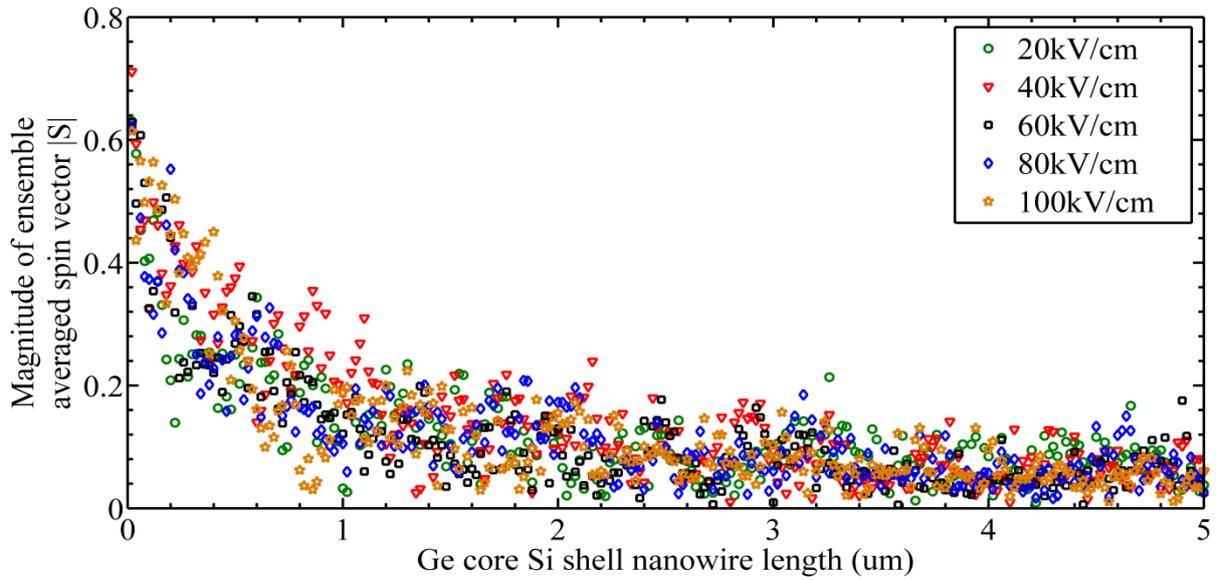

Fig.9. Variation in magnitude of ensemble averaged spin vector $|S|$ with varying electrical field in Ge-Core/Si-Shell nanowire

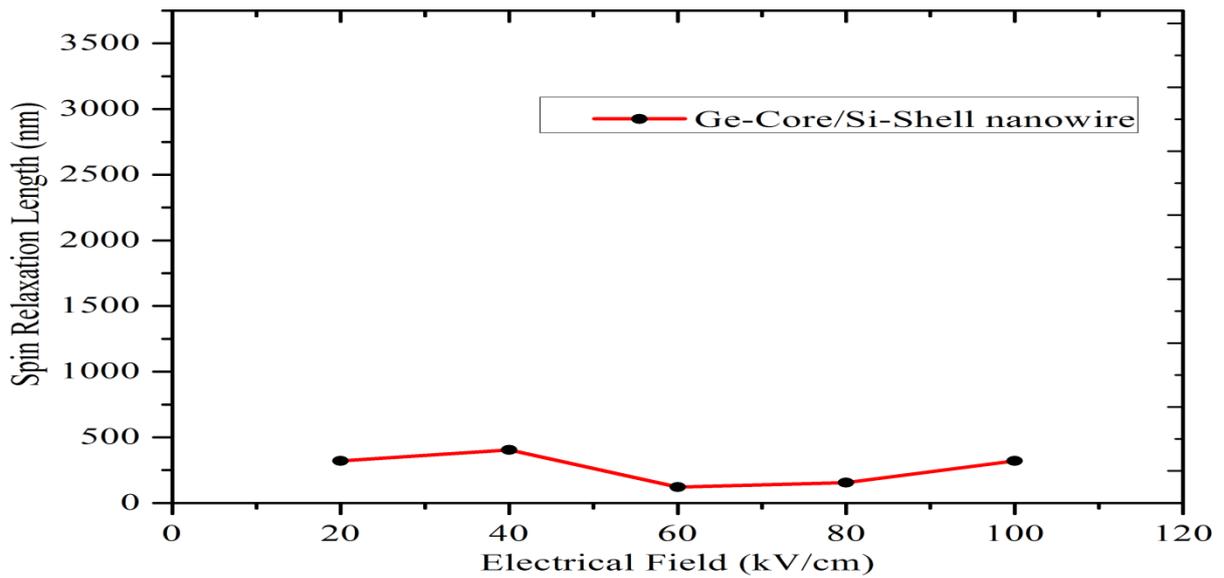

Fig.10. Variation in spin dephasing length with varying electrical field in Ge-Core/Si-Shell nanowire

C. Variation in magnitude of ensemble averaged spin vector $|S|$ and Spin dephasing length variation with Temperature in Si-Core/Ge-Shell and Ge-Core/Si-Shell nanowire:

Acoustic phonon scattering for Si-Core/Ge-Shell and Ge-Core/Si-Shell nanowire depends on temperature as shown in equation 10. Increase in temperature causes increase in acoustic phonon scattering which results in randomization of $k$ and $\Omega s$. As a result dephasing occurs more quickly and therefore spin dephasing length decreases with increase in temperature.

$$\Gamma_{nm}^{ac}(k_x) = \frac{E_{ac}^2 k_B T \sqrt{2m^*}}{\hbar^2 \rho v^2} D_{nm} \frac{(1+2\alpha\varepsilon_f)}{\sqrt{\varepsilon_f(1+\alpha\varepsilon_f)}} \Theta(\varepsilon_f) \qquad (10)$$

Where $v$ is the sound velocity, $\rho$ is the crystal density, $E_{ac}$ is the acoustic deformation potential and $\Theta$ is the Heaviside step function. $D_{nm}$ is overlap integral associated with electron phonon interaction.[31]

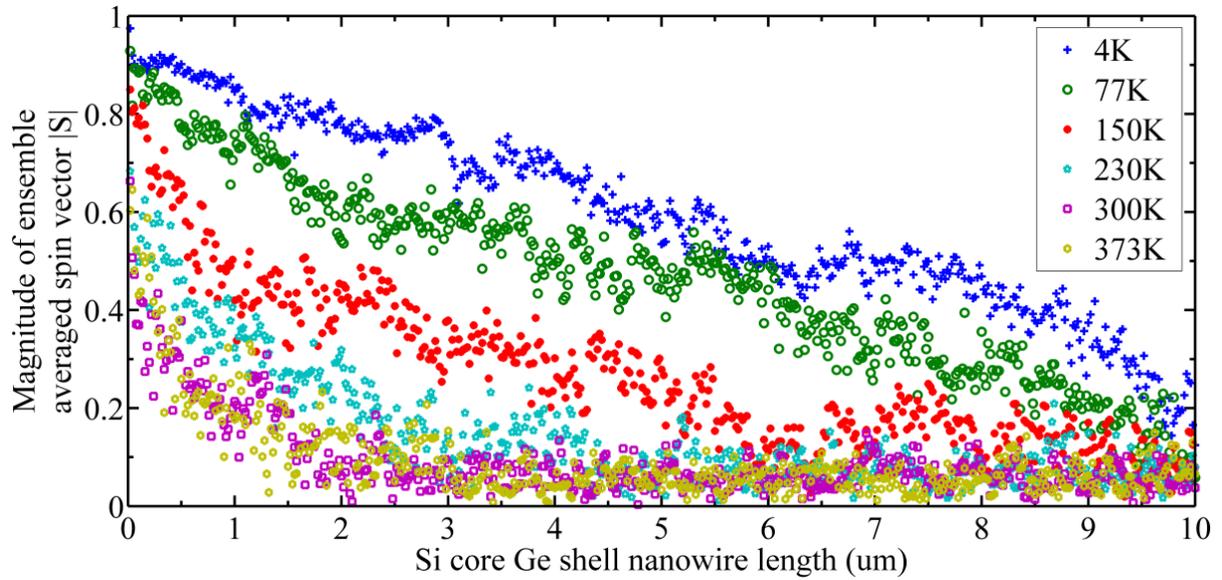

Fig.11. Variation in magnitude of ensemble averaged spin vector $|S|$ with varying temperature in Si-Core/Ge Shell nanowire

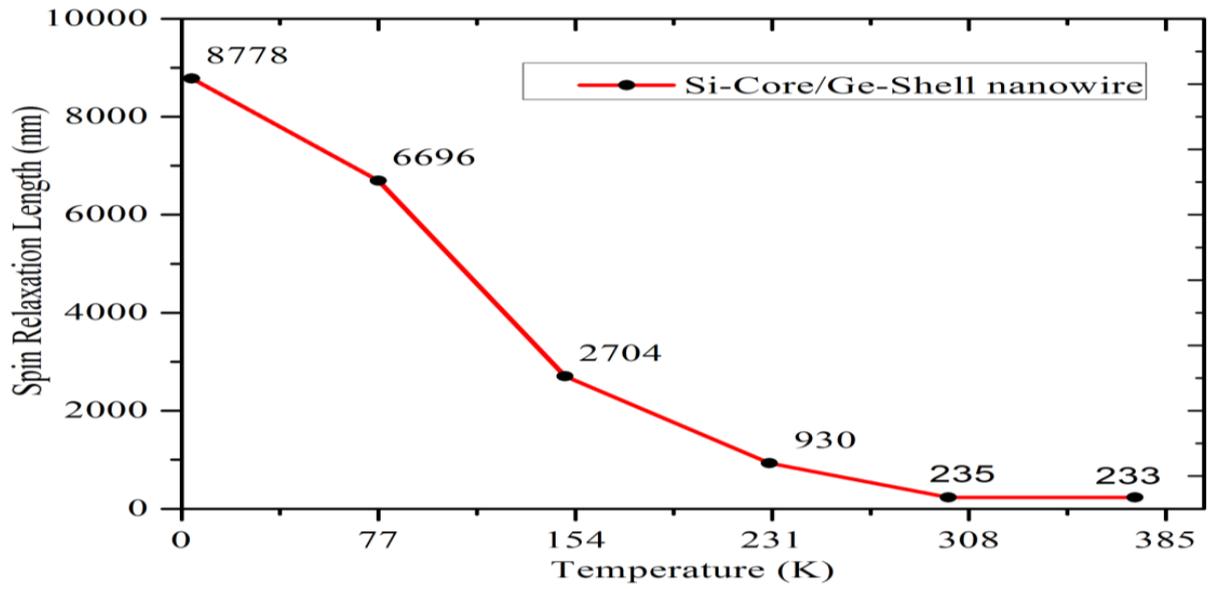

Fig.12. Variation in spin dephasing length with varying temperature in Si-Core/Ge-Shell nanowire

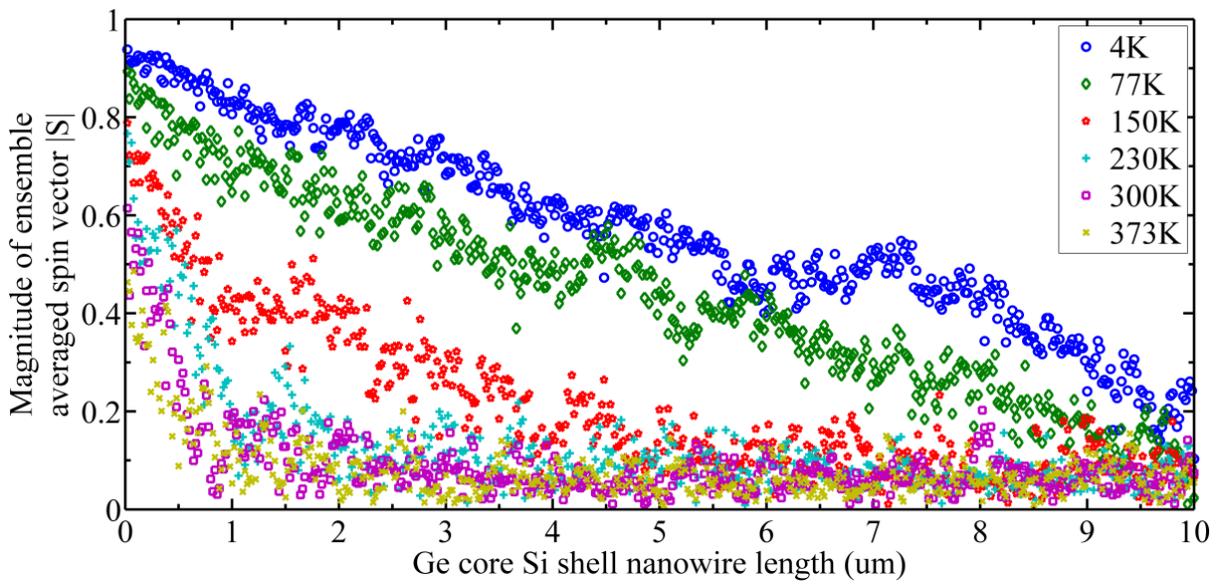

Fig.13. Variation in magnitude of ensemble averaged spin vector |$S$| with varying temperature in Ge-Core/Si-Shell nanowire

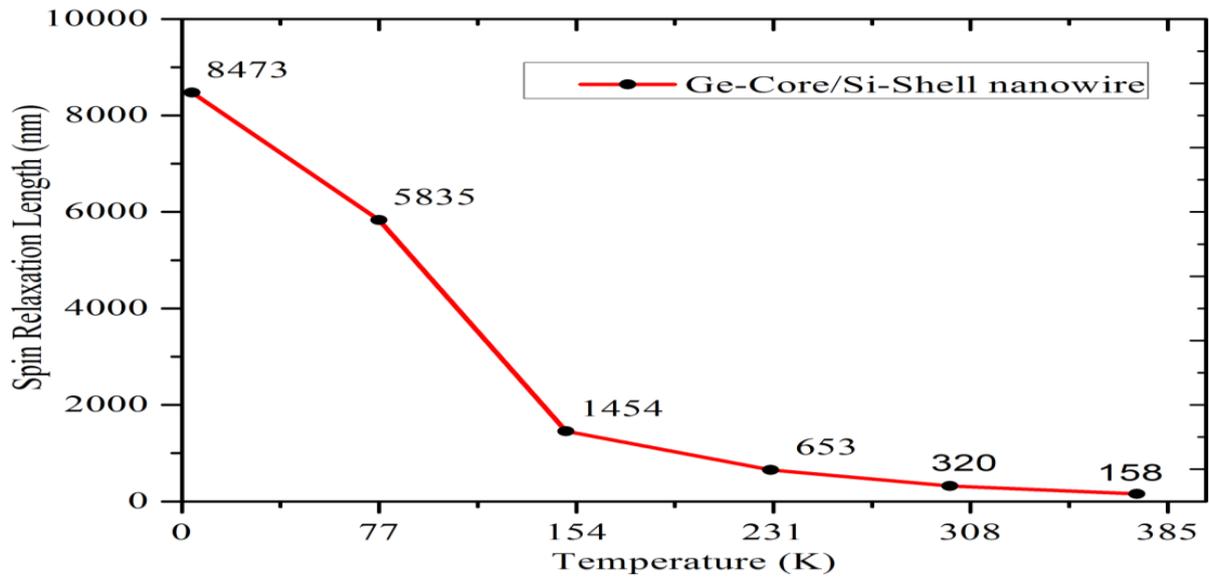

Fig.14. Variation in spin dephasing length with varying temperature in Ge-Core/Si-Shell nanowire

Table III shows the Spin dephasing length at various Temperature of Si-Core/Ge-Shell and Ge-Core/Si-Shell nanowire

TABLE III

| Temperature (K) | Spin Dephasing Length(nm) Ge-Core/Si-Shell nanowire | Spin Dephasing Length(nm) Si-Core/Ge-Shell nanowire |
|---|---|---|
| 4 | 8473 | 8778 |
| 77 | 5835 | 6696 |
| 150 | 1454 | 2704 |
| 230 | 653 | 930 |
| 300 | 320 | 235 |
| 373 | 158 | 233 |

Figures 11-14 show the variation in magnitude of ensemble averaged spin vector $|S|$ and spin dephasing length variation with Temperature for core diameter of 4nm at applied transverse electric field of 100 kV/cm. It clearly shows that the spin dephasing length decreases monotonically with increase in temperature.

# V. CONCULSION

In this article, we studied spin polarized electronic transport in Si-Core/Ge-Shell and Ge-Core/Si-Shell nanowire. Our study revealed that spin dephasing length depends on germanium core diameter in Ge-Core/Si-Shell and on silicon core diameter in Si-Core/Ge-Shell nanowire structure. We found that spin dephasing length increases with increase in Si-core diameter in Si-Core/Ge-Shell nanowire and spin dephasing length decreases with increase in Ge-core diameter in Ge-Core/Si-Shell nanowire. Effect of temperature on spin dephasing length is also investigated for core diameter of 4 nm. We found that Spin dephasing length is strongly dependent on temperature and it decreases monotonically with increase in temperature. Finally, we studied the variation in spin dephasing length with varying externally applied transverse electric field ranging from 20 kV/cm to100 kV/cm. In the electric field dependence study we found that spin dephasing length is weakly dependent upon applied electric field in Si-Core/Ge-Shell and Ge-Core/Si-Shell nanowire structure.

## ACKNOWLEDGEMENT

The authors thank the Department of Science and Technology of the Government of India for partially funding this work.